\documentclass[aps,prb,showpacs,twocolumn,groupedaddress,amsmath,amssymb]{revtex4-1}
\usepackage{graphicx}
\usepackage{bm}

\newcommand{\com}[2]{\ensuremath{[#1,\;#2]}}
\newcommand{\acom}[2]{\ensuremath{\{#1,\;#2\}}}
\newcommand{\abs}[1]{\ensuremath{|#1|}}
\newcommand{\tr}{\ensuremath{\operatorname{tr}}}
\newcommand{\ev}[1]{\ensuremath{\langle #1 \rangle}}
\newcommand{\sgn}{\ensuremath{\operatorname{sgn}}}

\newcommand{\id}{\ensuremath{\openone}}

\newcommand{\void}[1]{}

\begin{document}
	\title{Monitoring quantum transport: Backaction and measurement correlations}
	
	\author{Robert Hussein}
	\author{Jorge G\'omez-Garc\'ia}
	\author{Sigmund Kohler}
	\affiliation{Instituto de Ciencia de Materiales de Madrid, CSIC, Cantoblanco, E-28049 Madrid, Spain}
	\date{\today}
	
	\begin{abstract}
		We investigate a tunnel contact coupled to a double
		quantum dot (DQD) and employed as charge monitor for the latter.
		We consider both the classical limit and the quantum regime.  In the
		classical case, we derive measurement correlations from conditional
		probabilities, which yields quantitative statements about the parameter
		regime in which the detection scheme works well.  Moreover, we demonstrate that
		not only the DQD occupation but also the corresponding current may strongly
		correlate with the detector current.  The quantum mechanical solution,
		obtained with a Bloch-Redfield master equation, shows that the backaction of
		the measurement tends to localize the DQD electrons and, thus,
		significantly reduces the DQD current.  Moreover, it provides the effective
		parameters of the classical treatment.  It turns out that already the
		classical description is adequate for most operating regimes.
	\end{abstract}
	
	\pacs{
	73.23.Hk, %
	84.37.+q, %
	05.60.Gg %
	}
	\maketitle

	\section{Introduction}
	
	The conductance of a quantum point contact can be influenced significantly
	by the capacitive interaction with a close-by electron.  In this way it
	can act as detector for the charge state of a quantum
	dot in its vicinity.  After an early proof of principle,\cite{FieldPRL1993a}
	such a charge detector became a standard element in quantum dot design.
	Its practical use is to monitor single-electron tunneling through a quantum
	dot \cite{GustavssonPRL2006a, FujisawaScience2006a, FrickePRB2007a,
	IhnSSC2009a, GuttingerPRB2011a} and to measure charging diagrams with high
	precision.\cite{TaubertRSI2011a} Alternative detector concepts
	based on shifting a level across the Fermi energy of a lead
	\cite{WisemanPRB2001a} or tuning a DQD into and out of resonance
	\cite{KreisbeckPRB2010a} have been proposed as well.
	On the formal level, the measurement quality of such charge detection can
	be expressed by the correlation between the detector current and the dot
	occupation.\cite{KohlerEPJB2013a}
	
	In contrast to a single quantum dot, a DQD with strong inter-dot coupling
	possesses delocalized electron states which suffer from decoherence
	when their charge distribution is probed.  Such measurement backaction has
	been investigated theoretically for the readout of charge qubits
	\cite{WisemanPRB2001a, GiladPRL2006a, AshhabPRA2009a, KreisbeckPRB2010a} and the
	adiabatic passage of electrons.\cite{RechPRL2011a}  Typically a charge
	detector is strongly biased and, thus, entails non-equilibrium noise to the
	system to which it couples.  In this way it can induce
	pump currents \cite{KhrapaiPRL2006a, HusseinPRB2012a} and phonon-assisted
	tunneling.\cite{TaubertPRL2008a}  This complex interplay between measurement,
	decoherence, and non-equilibrium dynamics raises interest in correlations
	between the detector currents, the charge, and the current in a DQD.
	
\begin{figure}[b]
		\includegraphics[width=0.9\columnwidth]{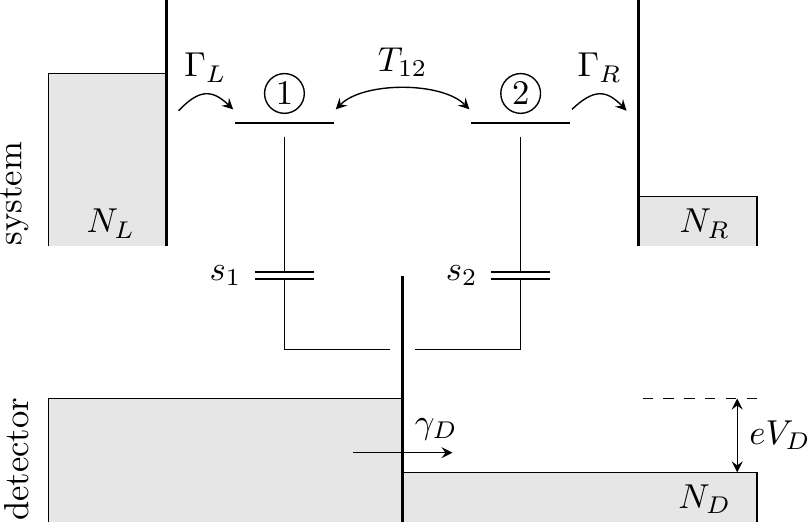}\caption{\label{fig:model}
		Quantum point contact in the tunnel regime acting as charge monitor for an
		undetuned but biased DQD.  Electrons on the latter increase the tunnel barrier and,
		thus, reduce the detector current.
		}
	\end{figure}

	In this work, we study a quantum point contact in the tunnel regime acting
	as charge monitor for a DQD, as is sketched in Fig.~\ref{fig:model}.
	Focusing on the correlations between the detector current and DQD
	observables, we reveal under which conditions the former correlates with
	both the charge and the current of the DQD.  In Sec.~\ref{sec:model}, we
	introduce a full quantum mechanical model for the DQD and the detector.
	For the specific calculations, we follow two different paths: First, in
	Sec.~\ref{sec.:classical_model} we consider the classical limit in which
	inter-dot tunneling is fully incoherent.  Hence, correlation functions can
	be expressed in terms of conditional probabilities.  For a full quantum
	mechanical treatment, we employ in Sec.~\ref{sec.:qm_model} a
	Bloch-Redfield master equation, which allows us to identify genuine quantum
	features such as decoherence and measurement backaction.  Comparing both
	limits provides the effective parameters and the limitations of the
	classical description.
	
	\section{Double quantum dot coupled to a charge detector}
	\label{sec:model}
	
	Our setup consists of a DQD formed
	by two single-level quantum dots in contact with electron source and
	drain.  Since double occupation of the DQD is inhibited by Coulomb
	repulsion, spin effects play a minor role and will be ignored.  This
	setup is described by the Hamiltonian $H_\text{DQD}+ H_\text{DQD,leads}$,
	where
	\begin{equation}
		\label{HDQD}
		H_\text{DQD} = T_{12}(c_1^\dagger c_2+c_2^\dagger c_1)
	\end{equation}
	models the DQD with vanishing onsite energies, tunnel coupling $T_{12}$,
	and the fermionic operators $c_\ell$, where $\ell=1,2$.  For ease of
	notation we use units with $e=\hbar=1$ and consider particle currents.  We
	assume a large Coulomb repulsion such that at most one electron can reside
	on the DQD, which means that the only energetically accessible states are
	the empty state $|0\rangle$ and the single-electron states $|\ell\rangle$.
	The coupling to the electron source and drain is given by
	\begin{equation}
		\label{Hleads}
		H_\text{DQD,leads} =
		\sum_{q,\alpha} V_{q\alpha} (c_{\ell_\alpha}^\dagger c_{q\alpha} + c_{q\alpha}^\dagger c_{\ell_\alpha})
		+\sum_{q,\alpha} \epsilon_q N_{q\alpha} ,
	\end{equation}
	where $c_{q\alpha}^\dagger$ are the fermionic creation operators for an
	electron in mode $q$ of lead $\alpha=L,R$ with the energy $\epsilon_q$.
	The mapping $\ell_\alpha$ takes the values $\ell_L=1$ and $\ell_R=2$, respectively.
	Tunneling between the DQD and the leads is determined by the spectral
	densities $\Gamma_\alpha(\epsilon) = (2\pi/\hbar)\sum_q |V_{q\alpha}|^2
	\delta(\epsilon-\epsilon_q) \equiv \Gamma_\alpha$ which we assume within a
	wideband limit energy independent.
	
	We restrict ourselves to a fully symmetric DQD with equal barrier
	capacitances. Then according to the Ramo-Shockley
	theorem,\cite{ShockleyJAP1938a, BlanterPR2000a} the displacement currents
	in the double dot circuit are such that the experimentally measured current
	is the average of the currents through the left and the right tunnel
	barrier, i.e., $I = \frac{1}{2}I_L - \frac{1}{2}I_R$.  Its noise spectrum
	defined below depends on the charge fluctuations of the DQD and
	reads\cite{BlanterPR2000a, MozyrskyPRB2002a,MozyrskyPRB2002b}
	\begin{equation}
		\label{CII}
		C_{II}(\omega) = \frac{1}{2}C_{I_LI_L}(\omega) + \frac{1}{2}C_{I_RI_R}(\omega)
		-\frac{\omega^2}{4} C_{NN}(\omega) .
	\end{equation}
	
	The charge detector is formed by a tunnel contact, see
	Fig.~\ref{fig:model}, and modeled by the Hamiltonian
	$H_D = \sum_k \epsilon_k c_k^\dagger c_k
	+\sum_{k'}\epsilon_{k'}c_{k'}^\dagger c_{k'}$ with the fermionic creation
	operators of the left and the right lead, $c_k^\dagger$ and
	$c_{k'}^\dagger$, respectively.  The tunnel coupling between the leads
	depends on the DQD occupation and reads\cite{GurvitzPRB1997a, BraggioJSM2009a,
	GolubevPRB2011a}
	\begin{equation}
		H_D^\text{tun} = (1-s_1 N_1-s_2 N_2)\sum_{kk'} t_{kk'}
		(c_k^\dagger c_{k'} + c_{k'}^\dagger c_k) \label{eq.:HD_tun},
	\end{equation}
	where $t_{kk'}$ denotes the tunnel matrix elements which we replace in a
	continuum limit by the conductance $G(\epsilon,\epsilon')
	= 2\pi \sum_{kk'} |t_{kk'}|^2 \delta(\epsilon-\epsilon_k)
	\delta(\epsilon-\epsilon_{k'}) \equiv G_D$ in units of $e^2/h$, which is
	also assumed energy independent. The number operators $N_\ell$
	in the prefactor reflect the fact that an electron on the
	DQD increases the potential barrier of the QPC and, thus, reduces the
	tunnel amplitudes.  The strength of this reduction depends on the
	interaction with the DQD which is quantified by the dimensionless
	parameters $s_1$ and $s_2$. For consistency, it must obey
	$s_1 N_1+s_2N_2\leq 1$ for all DQD occupations considered.
	
	\section{DQD in the classical limit}
	\label{sec.:classical_model}
	
	Within a classical approximation, we assume that the inter-dot tunneling is
	small such that $H_\text{DQD}$ practically commutes with the
	occupation operators $N_\ell$.  Then the DQD dynamics can be neglected for the
	computation of the tunnel rates.  Thus, we can adopt the golden-rule
	treatment of Ref.~\onlinecite{Ingold1992a} by which we obtain that an electron
	in state $k$ of the left lead may tunnel to state $k'$ of the right lead with
	probability $(2\pi/\hbar)|t_{kk'}|^2 \delta(\epsilon_k-\epsilon_{k'})
	(1-s_1 N_1 -s_2 N_2)^2$.  Expressing the probability for the
	initial many-body state in terms of Fermi functions and integrating over
	$\epsilon_k$ and $\epsilon_{k'}$, we obtain that for $N_1=N_2=0$ the QPC current can
	be described by a Poisson process with a rate $\gamma_0 = G_D\abs{V_D}$
	proportional to the bias voltage applied to the detector, $V_D$.
	\cite{Ingold1992a, BlanterPR2000a}  If an electron resides on the DQD,
	Coulomb repulsion reduces the tunnel rates according to $\gamma_0 \to
	\gamma \equiv \gamma_0(1-\tilde s_1 N_1-\tilde s_2 N_2)$, where $\tilde s_\ell =
	s_\ell(2-s_\ell)$ reflects the detector sensitivities.\cite{KohlerEPJB2013a}
	
	Subsuming these two cases, we can conclude that the QPC tunnel process inherits
	an additional randomness from the DQD occupation.  In more technical terms, the
	Poisson process turns into a Cox process with a rate
	\begin{equation}
		\label{gamma(t)}
		\gamma = \gamma_0(1-\tilde s_1 N_1-\tilde s_2 N_2)
	\end{equation}
	which depends on the transport process of the DQD.
	Thus, the average current through the detector, $\langle j\rangle$, can be
	expressed in terms of the DQD occupations.  While the same is true for the
	detector-DQD correlations, auto-correlations of the detector current
	contain also a (white) shot noise contribution, such that the power
	spectrum becomes \cite{BouzasAMM2006a}
	\begin{equation}
		\label{Cjj}
		C_{jj}(\omega) = \langle j\rangle
		+ \gamma_0^2 \sum_{\ell,\ell'=L,R} \tilde s_\ell\tilde s_{\ell'}C_{N_\ell N_{\ell'}}(\omega) .
	\end{equation}
	For an explicit derivation of the shot noise term,
	see Ref.~\onlinecite{MozyrskyPRB2002b}.  The fluctuations of the detector
	current are characterized by the frequency-dependent Fano factor $F(\omega)
	= C_{jj}(\omega)/\langle j\rangle$.  In consistency with
	Ref.~\onlinecite{KohlerEPJB2013a}, we find that good measurement
	correlations are accompanied by $F(\omega)\gg 1$, see
	Fig.~\ref{fig:cl:r(w)}(a) and discussion below.

	\subsection{Master equation for uni-directional transport}
	
	We consider a DQD with large bias such that electrons can enter exclusively from
	the left lead with tunnel rate $\Gamma_L$, while leaving to the right lead with
	tunnel rate $\Gamma_R$.  For our numerical study, we focus on a symmetric
	situation with $\Gamma_L=\Gamma_R\equiv\Gamma$.  Moreover, if the onsite
	energies of both dots are equal as well, inter-dot tunneling is direction
	independent with a rate $\Gamma_{12}\ll\Gamma$.  The restriction to small rates
	ensures that inter-dot tunneling is incoherent and, thus, consistent with the
	classical description.
	
	If at most one electron can reside in the DQD, we have to take into account
	the states $0$, $1$, and $2$, referring to an empty DQD and one electron in
	the left or the right dot, respectively.  Then the corresponding occupation
	probabilities obey the master equation $\dot P=\mathcal M P$, with
	\begin{equation}
		\label{M}
		\mathcal M=\begin{pmatrix} 
			-\Gamma_L &  0 &  \Gamma_R    \\
			\Gamma_L & -\Gamma_{12} &  \Gamma_{12}  \\
			0 &  \Gamma_{12} & -\Gamma_{12}-\Gamma_R
		\end{pmatrix}		,
	\end{equation}
	and $P = (P_0,P_1,P_2)^\mathsf{T}$, where $\mathsf T$ denotes transposition and
	$\ell=0,1,2$ labels the charge states of the DQD.  $P^\text{st}$ denotes
	with $\mathcal M P^\text{st}=0$ the stationary
	solution of the master equation.  Our central quantity for the
	computation of correlation functions is the conditional probability
	\begin{equation}
		\label{Pconditional}
		P(\ell,t|\ell',t') = [e^{\mathcal M(t-t')}]_{\ell\ell'} ,
	\end{equation}
	for the DQD being in state $\ell$ at time $t$ provided that it was in state
	$\ell'$ at the earlier time $t'<t$.  It is equivalent to the propagator of the
	master equation\cite{Risken1989a} and obeys $P(\ell,t+dt|\ell',t) = \delta_{\ell\ell'}+
	\mathcal{M}_{\ell\ell'} dt$.

	\subsection{DQD-detector correlations}
	
	The correlation of any DQD variable $X$ with the detector current $j$ can be
	obtained from the stochastic part of the rate $\gamma$ given by
	Eq.~\eqref{gamma(t)} and reads
	\begin{equation}
		\label{Cjx}
		C_{jX} = -\gamma_0 (\tilde s_1 C_{N_1X} +\tilde s_2 C_{N_2X}) .
	\end{equation}
	Since we are interested in the degree of correlation rather than in
	absolute values, we focus on the normalized correlation at a given
	measurement frequency $\omega$ which we define as
	\begin{equation}
		\label{rab}
		r_{ab}(\omega) =
		\frac{C_{ab}(\omega)}{\sqrt{C_{aa}(\omega)\,C_{bb}(\omega)}}\,.
	\end{equation}
	Its absolute value is a figure of merit for the detection quality and
	in the ideal case is of order unity.  In turn, for $|r_{jX}|\ll1$, the
	detector current is practically independent of $X$.
	
	In order to quantify the detection of the charge in dot $\ell=1,2$, we
	consider the correlation coefficient $r_{j N_\ell}$.  According to
	Eqs.~\eqref{Cjj} and \eqref{Cjx}, it can be expressed in terms of the DQD
	correlation functions of the populations which in the time domain read
	\begin{equation}
		\label{cl:CNN}
		C_{N_\ell N_{\ell'}}(t-t') = \langle N_\ell(t) N_{\ell'}(t')\rangle
		- \langle N_\ell\rangle\langle N_{\ell'}\rangle .
	\end{equation}
	Since $N_\ell$ can assume only the values 0 and 1, the first term on the
	right-hand side is given by the joint probability $P(\ell,t;\ell',t')$ for the
	DQD being in the states $\ell$ and $\ell'$ at the respective times.  Bayes'
	theorem relates this joint probability in the stationary limit to
	$P^\text{st}$ and the conditional probability \eqref{Pconditional} so that
	we obtain for $t\geq t'$ the expression
	\begin{equation}
		C_{N_\ell N_{\ell'}}(t-t')
		= P(\ell,t|\ell't') P_{\ell'}^\text{st} -P_\ell^\text{st}P_{\ell'}^\text{st} ,
	\end{equation}
	while the opposite time ordering $t<t'$ follows by relabeling.
	
	In order to obtain the correlation of the detector current with the DQD
	currents, we use Eq.~\eqref{Cjx} to write the detector current in terms of
	the DQD occupations and obtain $\langle N_1(t)I_R(t')\rangle$, as well as
	similar expressions with other combination of the indices $1$, $2$ and $L$,
	$R$.  Following Refs.~\onlinecite{KorotkovPRB1994a, KohlerEPJB2013a} we
	define the differential $dN_R(t) = I_R(t)dt$ which describes the change of
	the charge state in the right dot by a current flow to the leads.  Then we
	express the probabilities of all trajectories that contribute to $\langle
	N_1(t)I_R(t')\rangle dt = \langle N_1(t) dN_R(t')\rangle$ by the
	conditional probability \eqref{Pconditional}.
	
	For $t'<t$, the only contribution to the mentioned term stems from a
	trajectory starting at time $t'$ with an electron on the right dot which
	leaves during the infinitesimal time $dt$ to the right lead, such that the
	DQD will be in state $0$.  At a later time $t$, the left lead must be
	occupied.  For the opposite time ordering, the DQD starts at time $t'$ in
	state $1$, propagates to state $2$ at time $t$, while then an electron
	leaves to the right dot during $dt$.  The joint probability for these
	events reads
	\begin{align}
	\label{cl:CjI}
	&\ev{dN_R(t) N_1(t')}\nonumber\\
			&=\begin{cases}
		P(0,t+dt|2,t) P(2,t|1,t') P_1^\text{st}, &  t>t' \\
		P(1,t'|0,t) P(0,t+dt|2,t) P_2^\text{st}, &  t<t'
			\end{cases}.
	\end{align}
	
	The auto-correlation function of the DQD current at the right barrier
	requires an initial occupation of state $2$ at $t'$, tunneling to the right
	lead during $dt$, propagation from $0$ to $2$ during $t-t'$, and finally
	electron tunneling to the right lead.  This happens with probability
	\begin{equation}
		\label{cl:CII}
		\begin{split}
			\langle dN_R(t) dN_R(t')\rangle
			={}& P(0,t+dt|2,t)P(2,t|0,t')
			\\ & \times P(0,t'+dt|2,t') P_2^\text{st} ,
		\end{split}
	\end{equation}
	valid for $t'<t$ while the opposite time ordering again follows by
	relabeling.  At equal times, we have to add the shot noise
	contribution to obtain $C_{I_RI_R}(t-t') = \langle dN_R(t)
	dN_R(t')\rangle/dt^2 + \langle I_R\rangle\delta(t-t')$.  A derivation of the
	shot noise in the spirit of the present calculation can be found in
	Ref.~\onlinecite{KorotkovPRB1994a}.
	
	The correlation functions for all other possible combinations of the
	indices $L$ and $R$ can be obtained in the same way and are listed in
	Appendix \ref{app:correlations}.
	
	\begin{figure}[b]
		\centering
		\includegraphics{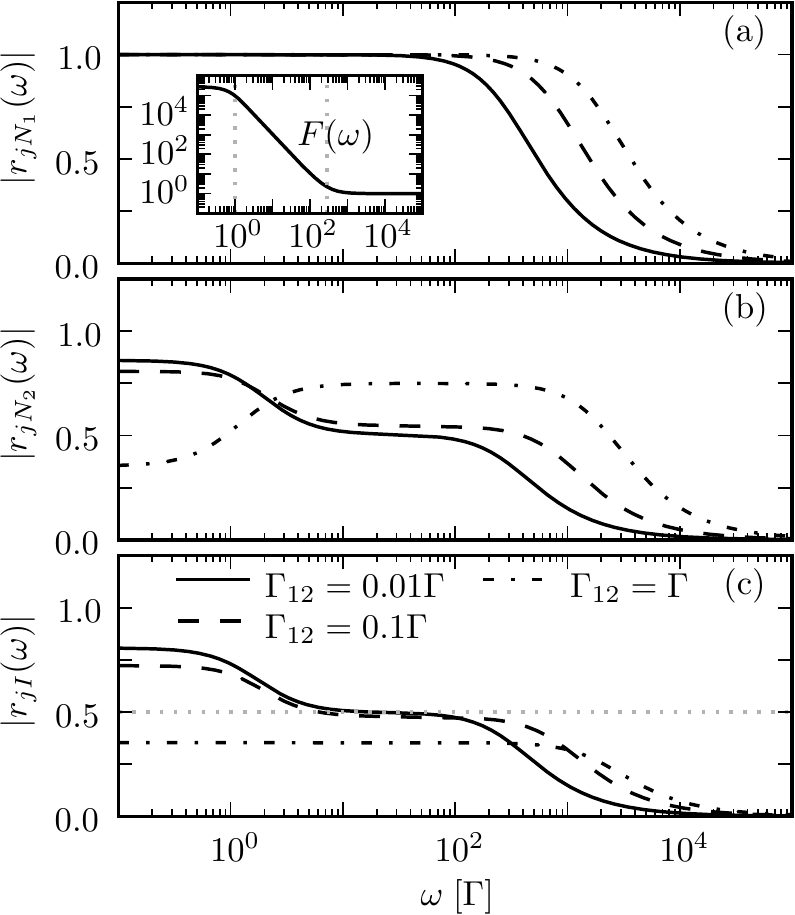}\caption{Classical frequency-dependent correlation coefficients between the detector
		current and (a) the occupation of the left dot, (b)
		the occupation of the right dot, and (c) the
		symmetrized current through the DQD for various
		inter-dot rates $\Gamma_{12}$.  The detector is characterized by a bare rate
		$\gamma_{0}=10^{8}\Gamma$ and the sensitivities $\tilde s_1=0.2$ and $\tilde
		s_2=0$, i.e., it couples to only the left dot. The inset in panel (a) shows 
		the frequency dependent Fano factor of the detector current for
		$\Gamma_{12}=0.01\:\Gamma$, where the dashed lines mark the crossover
		region between the plateaus.  The horizontal line
		in panel (c) marks the upper limit $1/2$ discussed in the text.
		}
		\label{fig:cl:r(w)}
	\end{figure}
	\begin{figure*}[tb]
		\centering
		\includegraphics{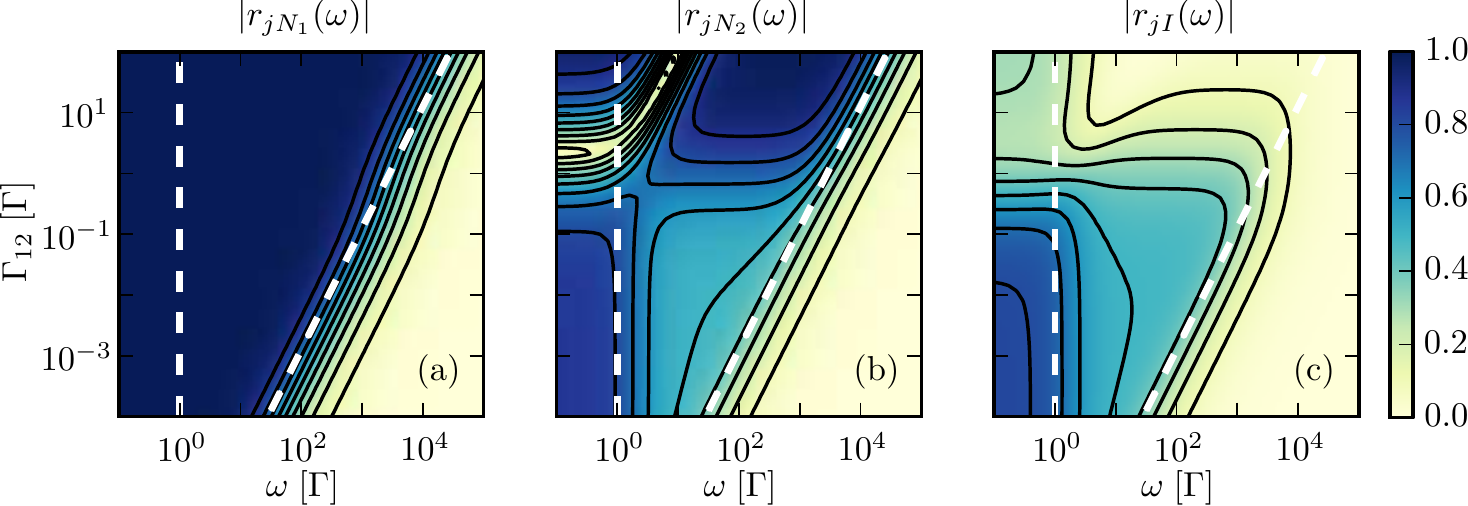}\caption{Classical correlation coefficients depicted in
		Fig.~\ref{fig:cl:r(w)} as function of the frequency $\omega$ and the
		inter-dot rate $\Gamma_{12}$ while all other parameters are the same. The
		dashed lines at $\omega=\Gamma$ and at $\omega = \omega_\text{max}$ mark
		the crossover between the different regimes discussed in the text.}
		\label{fig:cl:r(w,G)}
	\end{figure*}
	
	\subsection{Numerical results}
	
	For the numerical evaluation of the correlation coefficients, we diagonalize
	the matrix \eqref{M} to obtain a bi-orthonormal set of left and right
	eigenvectors, $u_i^\mathsf{T}$ and $v_i$, as well as the eigenvalues
	$-\lambda_i$, so that $\exp(\mathcal{M}t) = \sum_i v_i u_i^\mathsf{T}
	e^{-\lambda_i t}$ for $t>0$.  In doing so, we obtain for each correlation
	function a sum of decaying exponentials and a formal expression for the
	Fourier transformed of the propagator $P(\ell,t|\ell',0)$.  We restrict ourselves to a
	symmetric DQD with $\Gamma_L = \Gamma_R \equiv \Gamma$.  Then the latter
	parameter determines the frequency scale of the DQD dynamics.
	
	Investigating various correlation functions, we found that one has to
	distinguish three frequency regimes which can be characterized by the
	frequency-dependent Fano factor of the detector current derived in
	Appendix~\ref{app:clME} and depicted in the inset of
	Fig.~\ref{fig:cl:r(w)}(a).  First, if the measurement frequency is
	small, $\omega\lesssim\Gamma$, all correlation functions assume their
	zero-frequency value.  Typically the detector Fano factor is
	several orders of magnitude above the shot noise level, where its precise
	value depends much on the coupling strengths $\tilde s_1$ and $\tilde s_2$.
	Such low measurement frequencies correspond to static DQD properties, i.e.,
	time-averaged expectation values.  The crossover to the high-frequency
	limit occurs at
	\begin{equation}
		\omega_\text{max} = (2\gamma_0\Gamma_{12})^{1/2}\max(\tilde s_1,\tilde s_2) ,
		\label{wmax}
	\end{equation}
	which reflects the largest relevant frequency.  For
	$\omega\gtrsim\omega_\text{max}$,
	the Fano factor assumes the Poissonian value $F(\omega)\approx 1$ while all
	DQD-detector correlations practically vanish.  Thus, on such large frequency
	scales and on the corresponding short time scales, the detector cannot provide
	information about the DQD.  The proportionality of the upper limit,
	$\omega_\text{max}\propto \gamma_0^{1/2}$, has also been found for
	detecting the charge of a single quantum dot.\cite{KohlerEPJB2013a}  In the
	intermediate regime, we find $F(\omega)\propto\omega^{-2}$.  There the
	correlation coefficients provide information about the possibility of
	time-resolved measurement.  This generic global behavior of the
	detector-DQD correlations relates the possibility of charge detection to
	the emergence of super-Poissonian detector noise.  Physically, this reflects
	switching between two values of the detector current and the associated
	bunching of the electrons flowing through the detector.
	
	\subsubsection{Charge detection}
	
	Figures \ref{fig:cl:r(w)} and \ref{fig:cl:r(w,G)} show the correlations
	between the detector current $j$ and the DQD for the coupling to only the
	left quantum dot.  Then for frequencies below $\omega_\text{max}$, the
	measurement correlation with the occupation of the left dot (panel a in
	both figures) assumes the ideal value $r_{jN_1}=1$.  This indicates the
	possibility of time-resolved detection of the charge on the left dot, as
	long as the Fano factor stays significantly above the shot
	noise.  Thus the time resolution of the present charge detection scheme is
	determined by Eq.~\eqref{wmax}.
	
	Since an electron on the right quantum dot originates from the left lead,
	it must have occupied the left dot at some earlier stage.  Then one
	naturally expects some remnant correlation between the occupations of both
	dots.  As a consequence, the detector not only correlates with the dot to
	which it couples, but also with the other dot as can be appreciated in
	Figs.~\ref{fig:cl:r(w)}(b) and \ref{fig:cl:r(w,G)}(b).  In the regime of
	weak inter-dot tunneling, $\Gamma_{12}\lesssim\Gamma$, this correlation
	decays as a function of $\omega$ via an intermediate plateau limited by the
	crossover frequencies $\Gamma$ and $\omega_\text{max}$.  For
	$\omega\lesssim\Gamma$, we find $r_{jN_2}\approx0.8$, i.e., the average
	population of dot~2 can be noticed to some extent.  In the intermediate regime,
	$\Gamma \lesssim\omega \lesssim\omega_\text{max}$, the correlation
	coefficient drops down to a value $1/2$.  Interestingly enough, for a
	strong inter-dot rate $\Gamma_{12} \gtrsim\Gamma$, the intermediate plateau
	becomes larger than in the zero-frequency limit, but the correlation
	coefficient always stays clearly below unity.
	
	In a realistic setup, a charge detector at a DQD is sensitive not only to
	the closer dot, but to some extent also to the other dot.  This raises the
	question whether the influence of the latter affects the measurement
	quality.  Being interested in time-resolved measurement, we focus on the
	intermediate frequency regime.  Figure~\ref{fig:cl:r(s)}(a) shows
	the correlation coefficient of the detector current with the occupation of
	the left dot as a function of both sensitivities.  It demonstrates that
	(almost) perfect correlation requires $\tilde s_1\gtrsim 2\tilde s_2$,
	i.e., the left dot must couple at least twice as strong as the right dot.
	An extreme case of very small correlation is found for $\tilde
	s_2\approx2\tilde s_1$.  There the behavior is even counter-intuitive since
	reducing further the coupling to the left dot increases the correlation.
	Finally, the correlation with the population of the right dot (not shown)
	behaves accordingly.  It can be obtained by interchanging the labels 1 and
	2, which is non-trivial since reflection symmetry is absent owing to the
	bias voltage applied to the DQD.

	\subsubsection{Current detection}
	
	\begin{figure*}[tb]
		\includegraphics{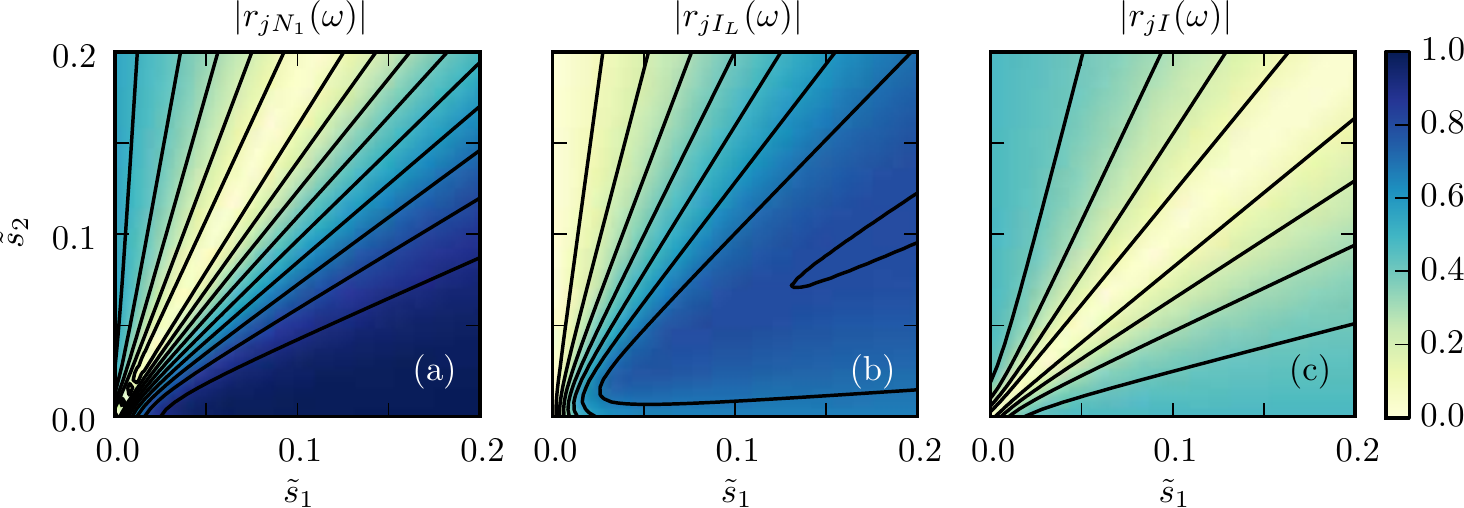}\caption{Classical correlation coefficients between the detector and (a)
		the occupation of the left dot, (b) the current through the left DQD
		barrier, and (c) the Ramo-Shockley current as a function of the detector
		sensitivities $\tilde s_1$ and $\tilde s_2$.  The frequency
		$\omega=50\Gamma$ corresponds to the middle of the second plateau at which
		time-resolved measurement is possible.  The inter-dot rate is
		$\Gamma_{12}=0.1\Gamma$, while all other parameters are as in
		Fig.~\ref{fig:cl:r(w)}.
		}
		\label{fig:cl:r(s)}
	\end{figure*}
	Even though the detector couples to the charge degree of freedom, it has
	been employed to reconstruct the corresponding time-resolved current
	\cite{GustavssonPRL2006a, FrickePRB2007a} and its full-counting
	statistics.\cite{FlindtPNASUSA2009a} A later theoretical investigation
	\cite{KohlerEPJB2013a} revealed that nevertheless the correlation
	coefficient between the detector current and the measured current is
	significantly smaller than unity.  Thus, knowledge about the transport
	mechanism must provides missing information.
	
	Figure \ref{fig:cl:r(s)}(b) depicts the correlation of the detector current
	with the current entering the DQD from the left lead at an intermediate
	measurement frequency.  It assumes its maximum $r_{jI_L}\approx 0.8$ for
	$\tilde s_1\approx 2\tilde s_2$.  Surprisingly, this value is slightly
	above the limit of $\sqrt{1/2}$ found for a detector coupled to a single
	quantum dot.\cite{KohlerEPJB2013a}
	A remarkable difference to the single quantum dot is also found for the
	Ramo-Shockley current $I=\frac{1}{2}I_L-\frac{1}{2}I_R$ which for a
	symmetric single quantum dot is fully uncorrelated with the detector
	current.\cite{KohlerEPJB2013a}  Figure~\ref{fig:cl:r(s)}(c), by contrast,
	reveals that this is not the case for a DQD unless both dots couple
	equally strongly to the detector.  If one coupling dominates, the
	correlation can be up to $r_{jI}=1/2$.

	\section{\label{sec.:qm_model}Quantum mechanical description}
	In order to obtain the quantum mechanical detector-DQD correlations, we
	employ a Bloch-Redfield master equation \cite{RedfieldIBMJRD1957a}
	augmented by a counting variable \cite{BagretsPRB2003a} that allows the
	computation of higher-order moments.  Since this master equation is
	Markovian, one can compute two-time expectation values of system variables via the quantum
	regression theorem.\cite{LaxPR1963a, LaxPR1967a, Carmichael2002a} In
	contrast to previous applications of this approach to quantum transport,
	\cite{EmaryPRB2007a, MarcosNJP2010a, UbbelohdeNC2012a} the detector current
	operator is not a usual ``electron jump term'' between the system and a
	lead, which requires a generalization of the formalism.  For the derivation
	we follow Ref.~\onlinecite{HusseinPRB2014a}.
	
	\subsection{Bloch-Redfield master equation}
	
	We start from the full DQD-lead-detector Hamiltonian
	and separate it into the DQD contribution $H_S$ given
	by Eq.~\eqref{HDQD}, the lead terms, and the tunneling from and to the
	four leads are involved.  After transforming the Liouville-von Neumann equation for the
	total density operator into the interaction picture with respect to the
	local terms, we derive a master equation that captures the lead tunneling to
	second-order.  Next, we multiply the full density operator with the
	$e^{i\bm\chi\bm N}$, where ${\bm \chi}=\{\chi_L,\chi_R,\chi_D\}$ contains the
	counting variables for the left and the right lead of the DQD and for the
	right lead of the detector, respectively.  Notice that owing to charge
	conservation, one counting variable for the detector is sufficient.  The
	vector $\bm N$ contains the corresponding lead electron numbers.
	By tracing out the leads, we obtain the master equation $\dot\rho =
	\mathcal L({\bm\chi})\rho$ with the generalized Liouvillian $\mathcal L({\bm\chi})\rho
	=-i\com{H_S}{\rho} + \mathcal L_S({\bm\chi})\rho +\mathcal L_D({\bm\chi})\rho$.
	The generalized density operator $\rho$ relates to the
	moment-generating function for the lead electrons as $\tr\rho = 
	\ev{e^{i\bm\chi\bm N}}$.  For DQD-lead tunneling in the large-bias
	limit, we find
	\begin{align}
	\mathcal L_S({\bm\chi})\rho={}&\Gamma_L \big[D(c^\dag_L)\rho + (e^{-i\chi_L}-1) c^\dag_L \rho c_L\big]\nonumber\\
	&+\Gamma_R \big[D(c_R)\rho+(e^{i\chi_R}-1) c_R \rho c^\dag_R\big]
	\end{align}
	with the Lindblad operator $D(x)\rho=x\rho x^\dag-\acom{x^\dag x}{\rho}/2$.
	
	The tunnel Hamiltonian of the detector, Eq.~\eqref{eq.:HD_tun}, contains
	besides lead terms the system operator $X \equiv \id-\sum_\ell s_\ell
	N_\ell$ which determines the generalized detector Liouvillian
	\begin{align}
\mathcal L_D(\bm\chi)\rho ={}&
Y_{-}(\bm\chi)\rho X +X\rho Y_{+}(\bm\chi) - XY_{-}\rho - \rho Y_{+} X,
	\end{align}
	where in the zero-temperature limit and for large bias voltage, $\abs{V_D}>2\abs{T_{12}}$,
	\begin{align}
\label{Y}
	Y_{\pm}(\bm\chi) 
	&= e^{i\chi_D\sgn V_D}\frac{G_D}{2} \big(
			\abs{V_D}X \pm\com{H_S}{X}
		\big).
	\end{align}
	The commutator in Eq.~\eqref{Y} is proportional to the inter-dot tunnel
	amplitude $T_{12}$ and, thus, can be neglected for large detector bias
	voltage, $\abs{V_D}\gg2\abs{T_{12}}$.  Therefore, we proceed with
	\begin{align}
\label{qm:LD}
	\mathcal L_D(\bm\chi)\rho
= { }	&\gamma_0 \big[D(X)\rho +(e^{i\chi_D\sgn V_D}-1) X\rho X\big],
	\end{align}
	where $\gamma_0 = G_D \abs{V_D}$ is the tunnel rate of the detector in the
	absence of the DQD.
	For typical parameters \cite{GustavssonPRL2006a, FrickePRB2007a,
	IhnSSC2009a} of $V_D=1\,\textrm{mV}$, $G_D\lesssim 0.1\, e^2/h$, DQD-lead
	rates $\Gamma$ of a few $\textrm{kHz}$, and inter-dot tunneling up to
	$T_{12}=100\,\mu\mathrm{eV}$, this corresponds to detector rates in the
	range of $10^5$--$10^8\,\Gamma$.

	\subsection{Charge and current correlations}
	\label{sec:cl:correlations}
	
	\begin{figure}[b]
		\centering
		\includegraphics{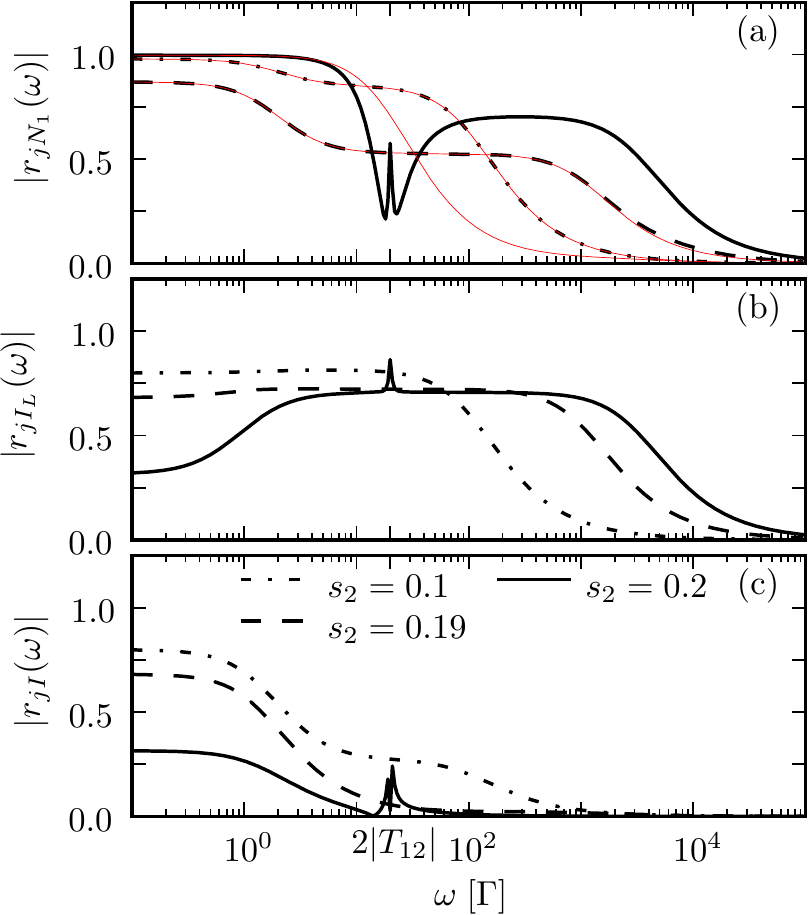}\caption{
		Quantum mechanical version of the correlation coefficients between the detector
		current and (a) the occupation of the left dot, (b)
		the current through the left DQD barrier, and (c) the
		symmetrized current through the DQD for various values of the sensitivity $s_2$
		while $s_1=0.2$ is fixed.
		The inter-dot tunnel coupling is $T_{12}=10\Gamma$, while
		all other parameters are as in Fig.~\ref{fig:cl:r(w)}. The thin red lines in
		panel (a) show the corresponding classical correlation coefficients for
		$\Gamma_{12}$ determined by the quantum-to-classical
		mapping in Eq.~\eqref{qm-cl-mapping}.
		}
		\label{fig:qm:r(w)}
	\end{figure}
	
	Also here we characterize the measurement by normalized correlation
	coefficients defined in Eq.~\eqref{rab}, but with the corresponding quantum
	mechanical expressions on the right-hand side.  Thus, we have to compute
	auto-correlations and cross correlations of the DQD occupations and the
	detector current.
	
	For the dot occupations, we define $C_{N_\ell N_{\ell'}}$ as in
	Eq.~\eqref{cl:CNN}.  From the quantum regression theorem \cite{LaxPR1963a,
	LaxPR1967a, Carmichael2002a} follows the frequency-dependent correlation
	function
	\begin{equation}
		\label{eq.:CNN}
		C_{N_\ell N_{\ell'}}(\omega) = 
		\tr[N_\ell\mathcal R(-i\omega)N_{\ell'}\rho_{\textrm{st}} + N_{\ell'}\mathcal R(i\omega)\rho_{\textrm{st}} N_\ell] .
	\end{equation}
	In order to formally perform the Fourier transformation, we have introduced
	the pseudoresolvent $\mathcal R(z) = \mathcal Q (z -\mathcal L)^{-1}
	\mathcal Q$ with $\mathcal Q=(\id-\rho_{\textrm{st}}\tr)$ the projector
	to the part of Liouville space orthogonal to the stationary state
	$\rho_{\textrm{st}}$ of the DQD.
	
	In contrast to the classical case, our master equation formalism allows us
	to treat all currents on equal footing, namely by computing derivatives of
	the generalized Liouvillian $\mathcal{L}(\bm\chi)$ with respect to the
	corresponding counting variable.  Proceeding as in
	Refs.~\onlinecite{EmaryPRB2007a, MarcosNJP2010a, UbbelohdeNC2012a}, we find
	\begin{equation}
		C_{I_\alpha I_\beta}(\omega)
		=\ev{\mathcal W_{\alpha\beta}}
		+\ev{\mathcal W_\alpha\mathcal R(-i\omega)\mathcal W_\beta}
		+\ev{\mathcal W_\beta\mathcal R(i\omega)\mathcal W_\alpha} \label{eq.:CII},
	\end{equation}
	where $\alpha,\beta\in\{L,R,D\}$ label the leads.  The
	superoperators $\mathcal W_{\alpha} =(\partial\mathcal L/\partial
	i\chi_\alpha)|_{\bm\chi=\bm0}$ and $\mathcal W_{\alpha\beta}
	=(\partial^2\mathcal L/\partial i\chi_\alpha\partial
	i\chi_\beta)|_{\bm\chi=\bm0}$ are Taylor coefficients of
	$\mathcal{L}(\bm\chi)$, where the first order provides the average currents
	$I_\alpha=\ev{\mathcal W_\alpha}$.  Formally, expression \eqref{eq.:CII} follows by
	substituting in Eq.~\eqref{eq.:CNN} the number operators by jump operators
	and adding the shot noise contribution $\ev{\mathcal W_{\alpha\beta}}$
	which vanishes unless $\alpha=\beta$.
	
	The frequency-dependent fluctuations~\eqref{CII} of the Ramo-Shockley
	current \cite{ShockleyJAP1938a, BlanterPR2000a}
	$I=\frac{1}{2}I_L-\frac{1}{2}I_R$ are linear combination of the above
	expressions.  They can also be obtained directly from the generalized
	density operator by transforming the counting
	fields according to \cite{MarcosNJP2010a}
	$\chi_L\to\chi_A+\chi_T/2$ and $\chi_R\to\chi_A-\chi_T/2$,
	where $\chi_T$ refers to the total current and $\chi_A$ accounts
	for temporary charge accumulation on the DQD.  Notice that we follow the
	sign convention of Ref.~\onlinecite{JinNJP2013a}, where the currents are
	positive when electrons flow from the lead to the DQD.
	
	For the cross correlations between currents and DQD occupations, we define the
	according expression
	\begin{equation}
		C_{I_\alpha N_{\ell}}(\omega)
		=	\tr[N_\ell\mathcal R(-i\omega)\mathcal W_\alpha\rho_{\textrm{st}}
		+\mathcal W_\alpha\mathcal R(i\omega)\rho_{\textrm{st}} N_\ell].
	\end{equation}
	Since we did not derive the latter correlation function in terms of a
	measurement procedure, it is an operationally defined quantity rather than
	an observable.
	For ease of notation we henceforth replace the subscript $I_D$ by $j$.

	\subsection{Classical limit of the quantum master equation}
	
	\begin{figure*}[tb]
		\includegraphics{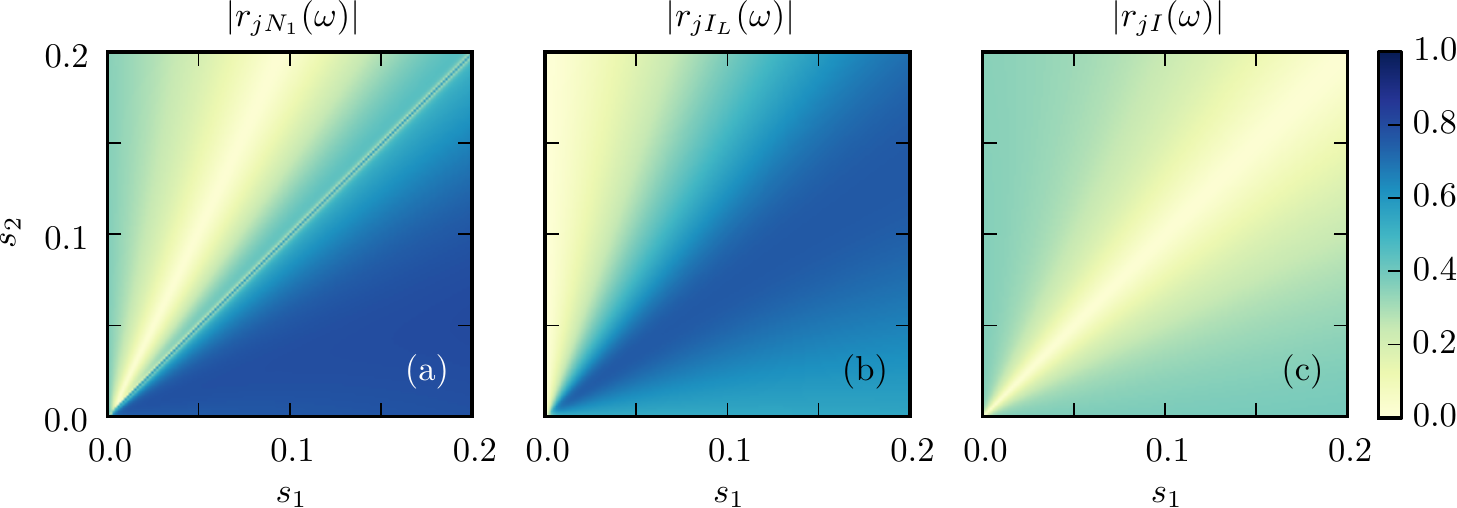}\caption{Quantum mechanical version of the correlation coefficients shown
		in Fig.~\ref{fig:cl:r(s)} as a function of the detector sensitivities
		$s_1$ and $s_2$. The frequency is $\omega=50\Gamma$, the tunnel rate is
		$T_{12}=10\Gamma$, while all other parameters are as in
		Fig.~\ref{fig:qm:r(w)}.  Notice that in the regime depicted, the classical
		and the quantum mechanical detector sensitivities relate as $s_\ell \approx
		\tilde s_\ell/2$.
		}
		\label{fig:qm:r(s)}
	\end{figure*}
	
	The classical limit of the Bloch-Redfield equation can be obtained by
	eliminating the coherences between the left and the right quantum dot
	in the limit of small inter-dot tunneling $T_{12}$.  This task is hampered
	by the fact that the natural basis of the Bloch-Redfield equation is given
	by the eigenstates of $H_S$ which, owing to the absence of a detuning, are
	always delocalized irrespective of how small $T_{12}$ is.  However, there
	exists a way out based on the comparison of the average currents in both
	limits.  While comparing the DQD currents yields an effective $\Gamma_{12}$
	in terms of $T_{12}$, the detector current provides a relation between the
	coupling strengths of the quantum mechanical model, $s_\ell$, and the
	classical couplings $\tilde s_\ell$.
	
	A straightforward computation of the stationary state of the classical
	master equation, see Eq.~\eqref{M}, yields the occupation numbers
	\begin{align}
\label{cl:N1}
\ev{N_1}&=\frac{\Gamma_L (\Gamma_R+ \Gamma_{12})}{\Gamma_L\Gamma_R +(2\Gamma_L+\Gamma_R) \Gamma_{12}} ,
\\
\label{cl:N2}
\ev{N_2}&=
\frac{\Gamma_L \Gamma_{12}}{\Gamma_L\Gamma_R +(2\Gamma_L+\Gamma_R) \Gamma_{12}} ,
	\end{align}	
	from which by use of $I_\text{DQD}^\text{cl} = \Gamma_R\langle N_2\rangle$
	immediately follows
	\begin{align}
	I_\text{DQD}^\text{cl}&=\frac{\Gamma_L\Gamma_R
\Gamma_{12}}{\Gamma_L\Gamma_R +(2\Gamma_L+\Gamma_R) \Gamma_{12}} .
	\end{align}
	Comparison with the corresponding expression for the quantum master
	equation, $\ev{\mathcal{W}_R}$, provides the effective classical
	inter-dot rate
	\begin{align}
\label{qm-cl-mapping}
	\Gamma_{12} = \frac{4 \abs{T_{12}}^2}{\Gamma_R + \gamma_0(s_1-s_2)^2}.
	\end{align}
	Obviously, the DQD current assumes its maximum for $s_1=s_2$, while
	it becomes much smaller when the two couplings are different (notice that
	typically $\gamma_0\ggg\Gamma_{L,R}$).  The reason for this current
	reduction is the fact that for $s_1\neq s_2$, the detector performs a
	position measurement of the DQD electrons.  Therefore, it destroys the
	coherence between the left dot and the right dot and, thus, forces the electron into
	the corresponding eigenstates, which leads to localization.  This localization
	is manifest in a current suppression which represents the main measurement
	backaction of the charge sensor to the DQD.  In the limiting case
	$s_1=s_2$, the measured quantity is the total electron number of the DQD
	which commutes with $H_S$ and therefore does not affect the coherences.
	Below we will find that in a large part of parameter space, the classical
	treatment with $\Gamma_{12}$ given by Eq.~\eqref{qm-cl-mapping} agrees very
	well with the full quantum mechanical solution.
	
	For the detector current, we insert the populations \eqref{cl:N1} and
	\eqref{cl:N2} into Eq.~\eqref{gamma(t)} to obtain
	\begin{align}
	j_\text{cl}&=G_D V_D(1-\tilde s_1\ev{N_1}-\tilde s_2\ev{N_2}),
	\end{align}	
	where the prefactor relates to the tunnel rate of the detector, $\gamma_0 =
	G_D \abs{V_D}$.  Comparison with the quantum mechanical expression and
	using the above result for $\Gamma_{12}$ provides a relation between the
	classical and the quantum mechanical detector sensitivities,
	\begin{equation}
		\tilde s_{\ell}=s_{\ell}(2-s_{\ell}),
	\end{equation}
	where for the small couplings considered in our numerical studies, $\tilde
	s_\ell\approx 2s_\ell$.

	\subsection{Numerical results}
	
	We already discussed above that the presence of the detector causes
	backaction which reduces the effective inter-dot rate
	$\Gamma_{12}$.  In particular, this rate becomes smaller with a larger
	difference between the two detector couplings, $|s_1-s_2|$.  Consequently,
	we expect that the frequency $\omega_\text{max}$ beyond which all
	DQD-detector correlations vanish [see Eq.~\eqref{wmax}] also
	depends on the coupling strength as well as on the bare detector rate
	$\gamma_0$.  The quantum mechanical correlation coefficients for different
	values of $s_2$ depicted in Fig.~\ref{fig:qm:r(w)} confirm this
	expectation.
	
	\subsubsection{Charge detection}
	
	Figures~\ref{fig:qm:r(w)}(a) and \ref{fig:qm:r(s)}(a) show the correlation
	coefficient between the occupation of the left quantum dot and the detector
	current. The former is compared with the classical result with the
	effective $\Gamma_{12}$ given by Eq.~\eqref{qm-cl-mapping}.  We find that
	as long as the two couplings are different, the values obtained from the
	quantum-to-classical mapping are practically the same as those of the
	quantum case.  A minor difference is visible at large frequencies for
	$s_2=0.19$.  Only when both couplings are equal, the quantum mechanical
	solution becomes rather different and is beyond the classical approach.  This
	corresponds to the situation discussed above in which the detector is
	sensitive to the total number of electrons on the DQD.  Then the
	DQD-detector Hamiltonian commutes with $H_S$ and, thus, it measures a good
	quantum number.
	
	This behavior is also found for the correlation for an intermediate
	frequency as a function of the couplings shown in
	Fig.~\ref{fig:qm:r(s)}(a).  The main difference to the corresponding
	classical solution (not shown) is found in a narrow region at $s_1=s_2$.
	As in the classical case, fulfilling the condition for good charge
	detection at dot~1, $r_{jN_1}\approx 1$, requires $s_1\gtrsim 2s_2$.
	
	\subsubsection{Current detection}
	
	Figure~\ref{fig:qm:r(w)} also shows the correlation coefficients with the
	DQD current through the left barrier (panel b) and with the Ramo-Shockley
	current (panel c).  Besides the global behavior already discussed for the
	correlation with the DQD occupations, we find for $s_1=s_2$ a sharp peak at
	a measurement frequency $2|T_{12}|$ which corresponds to the level
	splitting of the DQD.  As soon as both couplings differ minimally, this
	peak vanishes.  Since we consider $\gamma_0\ggg\Gamma$, a tiny difference
	of much less than one percent is already sufficient to suppress the peak.
	This demonstrates that the detector by and large destroys the quantum
	features of the DQD unless it couples to a good quantum number.
	
	Panels b and c of Fig.~\ref{fig:qm:r(s)} show the correlations of the
	detector and the DQD currents as a function of the sensitivities for an
	intermediate frequency.  It confirms the predictions from the classical
	treatment (cf.\ the corresponding panels of Fig.~\ref{fig:cl:r(s)}).  In
	particular, it shows that also quantum mechanically the current through the
	individual barriers may correlate strongly with the detector, while the
	Ramo-Shockley always correlates weakly.
	
	\section{Conclusions}
	
	We have studied a tunnel contact employed as charge sensor for a strongly
	biased DQD such that electrons are detected while being transported.  The
	central idea of this scheme is a capacitive coupling between the two
	subsystems by which electrons on the DQD reduce the transmission of the
	tunnel contact.  We have characterized this measurement by correlation
	coefficients of the detector current and DQD observables both in the
	classical limit and within a full quantum mechanical approach.  The
	comparison of these limits allowed us to investigate the backaction on the
	coherence of DQD electrons.
	
	A key ingredient to the classical description is a phenomenological
	incoherent inter-dot transition rate that enters the master equation for
	the DQD populations.  It determines the conditional probabilities of the
	DQD and, thus, the joint probabilities that enter the two-time correlations
	under investigation.  This approach represents a generalization of the one
	used for calculating current-current correlations \cite{KorotkovPRB1994a}
	and measurement correlations \cite{KohlerEPJB2013a} for a single electron
	transistor.
	
	The correlation coefficients studied provide a limiting
	frequency beyond which measurement is no longer possible and
	which determines the time-resolution of the detection scheme.  This limit
	increases with the detector rate, the inter-dot rate, and the detector
	sensitivity.  The possibility of charge detection depends also crucially on
	the ratio between the capacitive couplings to each dot: A charge on a
	particular dot can be monitored reliably only if it couples to the detector
	at least twice as strong as an electron on the other dot.
	With the time-resolved DQD populations at hand, one can reconstruct the
	corresponding time-dependent current, at least under the assumption of
	unidirectional transport.  This is reflected by a significant, but not
	perfect correlation between the detector and the DQD currents.  Rather
	surprisingly, it is slightly larger than for the corresponding setup with a
	single-electron transistor, despite the more complicated transport
	mechanism of the present case.
	
	On the quantum mechanical level, we used a method based on a Bloch-Redfield master
	equation augmented by a counting field.\cite{EmaryPRB2007a, MarcosNJP2010a}
	In order to capture also the detector current, we generalized this method
	to the presence of ``jump terms'' that do not alter the DQD occupation, but
	describe the detector current.
	A main issue for such quantum mechanical position measurement is its
	backaction to the coherence of the measured system. In the present case it
	is manifest in an additional localization of the DQD electrons.  On the one
	hand, this leads to a significant reduction of the DQD current, on the
	other hand, it pushes the system towards its classical limit.  Indeed our
	quantitative analysis revealed that the classical description is adequate
	whenever the detector current correlates strongly with one of the DQD
	occupations.  A natural expectation is that this tendency should be even
	stronger in the presence of couplings to external degrees of freedom such
	as the electronic circuitry or substrate phonons.
	
	Even though we restricted ourselves to a narrow part of parameter space, we
	observed a rather rich behavior.  Thus, a full understanding of the
	detection scheme may require to take further ingredients into account.
	Besides the already mentioned influence of external degrees of freedom, this
	could be a detuning which leads to additional localization.
	
	\begin{acknowledgments}
		This work was supported by the Spanish Ministry of Economy and Competitiveness
		through grant no.\ MAT2011-24331 and by a FPU scholarship (R.H.).
	\end{acknowledgments}
	
	\appendix
	
	\section{Conditional probabilities and correlation functions}
	\label{app:correlations}
	
	In Sec.~\ref{sec:cl:correlations}, we explicitly
	derived the correlation functions $C_{I_RN_1}$ and $C_{I_RI_R}$, see
	Eqs.~\eqref{cl:CjI} and \eqref{cl:CII}.  For completeness, we here sketch
	the derivation of all other expressions required for the evaluation of the
	correlation coefficients discussed in the main text.
	
	As discussed above, correlations between DQD occupations and DQD current
	from lead $\alpha=L,R$ can be expressed by the differential $dN_\alpha =
	I_\alpha dt$ and joint probabilities.  For the currents
	through the left and the right contact, we thus obtain
	\begin{align}
\langle I_L\rangle dt ={}& \langle dN_{L}(t)\rangle=P(1,t+dt|0,t)P_{0}^{\text{st}},
\label{eq:IL}
\\
\langle I_R\rangle dt ={}& \langle dN_{R}(t)\rangle=P(0,t+dt|2,t)P_{2}^{\text{st}},
\label{eq:IR}
	\end{align}
	where we find in consistency with charge conservation $\langle I_L\rangle =
	\langle I_R\rangle$.
	
	The two-time correlations follow from the conditional probability
	\eqref{Pconditional} and Bayes' theorem which yields
	\begin{align}
\label{eq:NLdNL}
&\ev{N_{1}(t)dN_{L}(t')}\\
		&=\begin{cases}
	P(1,t'+dt|0,t')P(0,t'|1,t)P_{1}^{\text{st}}, & t<t' \\
	P(1,t|1,t'+dt)P(1,t'+dt|0,t')P_{0}^{\text{st}}, & t>t'  
			\end{cases},
\nonumber
	\end{align}
	\begin{align}
\label{eq:NRdNL}
		&\langle N_{2}(t)dN_{L}(t')\rangle\\&=\begin{cases}
	P(1,t'+dt|0,t')P(0,t'|2,t)P_{2}^{\text{st}}, & t<t' \\
	P(2,t|1,t'+dt)P(1,t'+dt|0,t')P_{0}^{\text{st}}, & t>t' 
			\end{cases},     
\nonumber
	\end{align}
	\begin{align}
\label{eq:NRdNR}
		&\langle N_{2}(t)dN_{R}(t')\rangle\\&=\begin{cases}
	P(0,t'+dt|2,t')P(2,t'|2,t)P_{2}^{\text{st}}, & t<t'  \\
	P(2,t|0,t'+dt)P(0,t'+dt|2,t')P_{2}^{\text{st}}, & t>t' 
			\end{cases}.     
\nonumber
	\end{align}
	Subtracting $\langle N_\ell\rangle\langle dN_{\ell'}\rangle$ and dividing
	by $dt$ yields the desired occupation-current correlations.
	
	Accordingly, the correlation function of the left DQD current can be expressed
	in terms of
	\begin{align}
\label{eq:probcurrents}
&\langle dN_{L}(t)dN_{L}(t')\rangle\\
        &=P(1,t+dt|0,t)P(0,t|1,t'+dt)P(1,t'+dt|0,t')P_{0}^{\text{st}}.
\nonumber
	\end{align}
	Notice that this expression provides the auto-correlation function
	$C_{I_LI_L}(t-t')$ only for $t\neq t'$, while for equal times, the shot
	noise contribution $\langle I_L\rangle \delta(t-t')$ must be added.
	\cite{KorotkovPRB1994a}

	\section{\label{app:clME}Alternative solution of the classical model and
	Fano factor}
	
	The numerical method for computing the quantum mechanical correlation
	functions in Sec.~\ref{sec.:classical_model} can be employed as well
	for the classical limit.  The classical master equation is formally an
	equation of motion for the diagonal matrix elements of the density matrix
	in the localized basis.  Since in the DQD Liouvillian the dissipative term
	that describes the influence of the detector, see Eq.~\eqref{qm:LD}, is
	also diagonal in this basis, we merely have to replace the augmented
	Liouvillian $\mathcal{L}(\bm\chi)$ by
	\begin{align}
\label{app:Mchi}
\mathcal M({\bm\chi})
={ }&\mathcal{M}
     + \Gamma_L(e^{-i\chi_1}-1)\mathcal{J}_L 
     + \Gamma_R (e^{i\chi_2}-1)\mathcal{J}_R \nonumber\\
	&+\gamma_0 (e^{i\chi_D\sgn V_D}-1)\mathcal{J}_D
	\end{align}
	with $\mathcal{M}$ as in Eq.~\eqref{M} and the DQD-lead jump operators
	\begin{align}
\mathcal{J}_L =
			\begin{pmatrix}		 0 & 0 & 0 \\1 & 0& 0 \\ 0 & 0 & 0 \end{pmatrix}, \qquad
\mathcal{J}_R =
			\begin{pmatrix}		 0 & 0 & 1 \\0 & 0& 0 \\ 0 & 0 & 0 \end{pmatrix},
	\end{align}
	while the detector jump operator
	\begin{align}
\mathcal{J}_D =
		\begin{pmatrix}1 & 0 & 0\\0 & 1-\tilde s_1 & 0 \\ 0 & 0 & 1-\tilde s_2
		\end{pmatrix}
	\end{align}
	follows from Eq.~\eqref{qm:LD} by ignoring non-diagonal contributions.  We
	have confirmed all results of Sec.~\ref{sec.:classical_model} with this
	method.
	
	This alternative method is rather convenient for obtaining an analytical
	expression for the detector Fano factor in the classical limit.  For this
	purpose, we compute the pseudoresolvent of $\mathcal{M}$ so that we can
	directly evaluate the auto-correlation function of the detector,
	$C_{jj}(\omega)$.  Thus, with the average current $\ev{j}$, we find the
	frequency-dependent Fano factor $F_D(\omega) = C_{jj}(\omega)/\ev{j}$.  For
	$\Gamma_L=\Gamma_R \equiv\Gamma$ and small inter-dot rates
	$\Gamma_{12}\ll\Gamma$, it becomes
	\begin{align}
\label{app:FD}
	F_D(\omega) &= 1+ \frac{2\gamma_0\Gamma_{12}}{\Gamma^2+\omega^2}
	\frac{\frac{3\Gamma^2+\omega^2}{\Gamma^2+\omega^2}\tilde s_1(\tilde s_1-\tilde s_2)+\tilde s_2^2}{1-\tilde s_1}
	\end{align}
	and assumes rather large values in the zero-frequency limit $\omega\ll
	\Gamma$.  For $\omega \gtrsim\Gamma$, we approximate the last factor in
	Eq.~\eqref{app:FD} by $\max(\tilde s_1,\tilde s_2)$ and obtain $F_D(\omega)
	= 1+(\omega_{\textrm{max}}/\omega)^2$ with $\omega_{\textrm{max}}$ given by
	Eq.~\eqref{wmax}.

\end{document}